\documentclass[final,5p,twocolumn]{elsarticle}

\usepackage{graphicx}          
\usepackage[dvips]{epsfig}    

\usepackage{xcolor}

\usepackage{amssymb}
\usepackage{amsthm}

\usepackage{amsmath}          

\usepackage{algorithmic}

\journal{Mechatronics}

\begin{document}

\begin{frontmatter}

\title{Time-delay based output feedback control of fourth-order oscillatory systems} 

\author{Michael Ruderman}
\ead{michael.ruderman@uia.no}


\address{
Department of Engineering Sciences, University of Agder, P.B. 422,
Kristiansand, 4604, Norway}

\begin{abstract}                          
We consider a robust stabilization of the fourth-order oscillatory
systems with non-collocated output sensing. Worth recalling is
that the fourth-order systems are relatively common in
mechatronics as soon as there are two-mass or more generally
two-inertia dynamics with significant elasticities in the link. A
novel yet simple control method is introduced based on the
time-delayed output feedback. The delayed output feedback requires
only the oscillation frequency to be known and allows for a robust
control design that leads to cancelation of the resonance peak. We
use the stability margins to justify the transfer characteristics
and robustness of the time-delay control in frequency domain. The
main advantage of the proposed method over the other possible
lead-based loop-shaping strategies is that neither time
derivatives of the noisy output nor the implementation of transfer
functions with a numerator degree greater than zero are required
to deploy the controller. This comes in favor of practical
applications. An otherwise inherently instable
proportional-integral (PI) feedback of the non-collocated output
is shown to be stabilized by the proposed method. The control
developed and associated analysis are also confirmed by the
experimental results shown for the low damped two-mass oscillator
system with uncertainties.
\end{abstract}


\end{frontmatter}

\newtheorem{thm}{Theorem}
\newtheorem{lem}[thm]{Lemma}
\newtheorem{clr}{Corollary}
\newdefinition{rmk}{Remark}
\newproof{pf}{Proof}

\section{Introduction}
\label{sec:1}

Time delays in a feedback control loop are usually associated with
degradation of performance and robustness and, in worth case, with
destabilization due to the evoked phase 'deficit' of the loop
transfer function. At the same time, there are situations where
time delays are used as controller parameters, cf.
\cite{Michiels2013}. Several important classes of dynamic systems,
including different type of oscillators, cannot be stabilized by
static output feedback, although they might be stabilizeable by
inserting an artificial time delay into the feedback, cf. e.g.
\cite{fridman2016}. While for a well developed (Lyapunov based)
stability analysis of time-delay systems we refer to seminal
literature, see e.g. \cite{sipahi2011}, \cite{kharitonov2013},
\cite{fridman2014} and references therein, a purposeful use of a
time-delayed feedback for stabilization is a less studied topic in
the control applications. A former work \cite{abdallah1993} used a
positive delayed output feedback for stabilizing the second-order
oscillatory system. Later, the output feedback stabilization
problem of a chain of integrators using multiple delays was
addressed in \cite{niculescu2004}. More specifically, it was shown
in \cite{niculescu2004} that a chain of $n$ integrators can be
stabilized by a proportional plus delay controller including $n-1$
delays, or by a chain of $n$ delay blocks. As early as two decades
ago, it was already recognized that delay properties can be also
useful, since introducing delays voluntarily can benefit the
control, see an overview of some advances and open problems with
time-delay systems in \cite{richard2003}. Still, to the best of
the author's knowledge, a positive delayed feedback was used only
by the Pyragas control \cite{pyragas1992} for stabilization of
unstable periodic orbits of a chaotic system. Later, a similar
delay-based strategy was pursued in \cite{balanov2004} to control
the noise-induced oscillations in nonlinear second-order systems.
However, unlike the approach proposed in the present work, an
opposite sign of the difference between the delayed and current
state of the system and the output rate instead of the output
value itself were used in \cite{balanov2004} for feedback.

While most of the known works on time delay systems (some of which
were mentioned above) deal with stabilization of the
delay-affected plants and networks, the approach provided in the
present work is rather to use a purposefully injected delay as a
control parameter. The main motivating idea behind the proposed
control is the fact of anti-phase between the input $u$ and
oscillatory mode of a fourth-order system. Here, it is worth
recalling that the fourth-order systems are relatively common in
mechatronics as soon as there are two-mass or more generally
two-inertia dynamics with significant elasticities in the link.
Notable applications can be found, for example, with payloads
driven via the ropes or cables, such as elevators, cranes, winches
and others. It is also worth emphasized that the proposed method
uses only the output information $x$ but not its time derivatives
or other internal dynamic states, cf. with \cite{sipahi2011}. This
comes in favor of different practical applications like, for
example, with non-collocated sensing and actuation, or noisy
output sensing where the time derivatives are not available. One
of the main contributions of this work can be viewed in the
introduced control
\begin{equation}\label{eq:1:1}
u(t) = K_d \bigl(x(t)-x(t-\tau)\bigr),
\end{equation}
with the dedicated time-delay parameter $\tau$. We have to notice
that some preliminary results were reported in
\cite{ruderman2021}, where \eqref{eq:1:1} was shown for the first
time. In view of this, the claimed novelty and results of the
present work are:
\begin{itemize}
    \item[-] detailed analysis of the closed-loop with
    \eqref{eq:1:1};
    \item[-] design of the robust time-delay-based stabilizer for
    PI-controlled non-collocated fourth-order plants;
    \item[-] experimental confirmation of the effectiveness and performance of
    the oscillation compensator \eqref{eq:1:1}.
\end{itemize}
The rest of the paper is organized as follows. In Section
\ref{sec:2}, the problem formulation is given. The main results of
the introduced feedback control \eqref{eq:1:1}, its stability
analysis, and parameterizations are presented in section
\ref{sec:3}. The stabilization problem of a PI-feedback controlled
fourth-order plant with non-collocated sensing and actuation is
addressed in section \ref{sec:4}. This is followed by the
associated experimental example demonstrated in section
\ref{sec:5}. Brief summary and discussion are given in section
\ref{sec:6}.

\textbf{Notation.} Throughout the text, the uppercase italic Latin
letters denote the vectors and matrices of the appropriate
dimension, while their lowercase counterparts denote the
variables. The dynamic variables in time domain are with argument
$t$, and in Laplace domain with argument $s$. The italic Latin
letters with time argument are used for denoting the measured
signals. The Latin and Greek letters denote the constants and
parameters, while the control gains are denoted by $K$ with a
subindex. $I$ is an identity matrix of an appropriate dimension.
Unless other specified, $\omega$ is the angular frequency, $j$ is
the imaginary unit of a complex number, and $|\Omega|$ and $\angle
\Omega$ are the magnitude and phase of a complex function
$\Omega$.

\section{Problem formulation}
\label{sec:2}

\begin{figure}[!h]
\centering
\includegraphics[width=0.99\columnwidth]{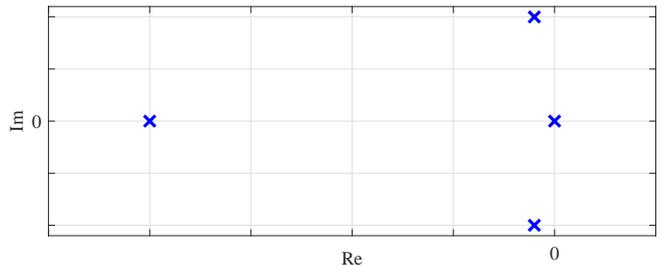}
\caption{Typical poles configuration of oscillatory system
\eqref{eq:2:1}, \eqref{eq:2:2}.} \label{fig:2:1}
\end{figure}

We consider a class of single-input-single-output (SISO)
oscillatory systems of the fourth-order with one free integrator.
Note that the integrative behavior of the system output is
purposefully required for elucidating the issue of a feedback
destabilization due to the lack of phase characteristics.
Otherwise, instead of the pole in origin, the systems considered
can also have an additional negative real pole. For the measurable
system input and output $u(t)$ and $x(t)$, respectively, the
transfer characteristics in Laplace domain are given by
\begin{equation}\label{eq:2:1}
G(s) = \frac{x(s)}{u(s)} = F (sI - A)^{-1}B.
\end{equation}
The exemplified system matrix and coupling vectors are
\begin{equation}\label{eq:2:2}
\nonumber
A = \left(%
\begin{array}{cccc}
-a_{1} &  -a_{2}  &    a_{3}      & -a_{4}  \\
1       & 0         &     0          & 0         \\
a_{5}  & a_{6}    &    -a_{7}     & -a_{8} \\
0       & 0         &    1           & 0
\end{array}%
\right), \quad B = \left(%
\begin{array}{c}
b \\
0      \\
0  \\
0
\end{array}%
\right),
\end{equation}
$F = (0, 0, 0, 1)$, respectively, with the constant parameters
$a_{1}, \ldots, a_{8}, b > 0$ which are satisfying $a_{3}, a_{5},
a_{7} \ll a_{1}, a_{2}, a_{4}, a_{6}, a_{8}$. A typical
configuration of the poles of the system \eqref{eq:2:1},
\eqref{eq:2:2} is exemplary shown in Figure \ref{fig:2:1}, without
numerical values.
Then, the associated fourth-order input-output transfer function
\eqref{eq:2:1} can be rewritten as
\begin{equation}\label{eq:2:3}
G(s) = \frac{N(s)}{D(s)} = \frac{k (s + z_1)}{s ( s + p_1) (s^2 +
2 \zeta \omega_0 s + \omega_0^2)}.
\end{equation}
Here the system parameters are accommodated in the gain factor
$k$, natural frequency $\omega_0$, damping ratio $\zeta$, and the
roots' coefficients of one real zero $z_1$ and one real pole
$p_1$, while $z_1 > \omega_0$; all coefficients are strictly
positive. Worth emphasizing that $0 < \zeta \ll 1$, while for
$\zeta = 1$ the conjugate complex pole pair collapses into the
double real pole, and the system \eqref{eq:2:3} becomes critically
damped without having to compensate for oscillatory behavior.

\begin{figure}[!h]
\centering
\includegraphics[width=0.99\columnwidth]{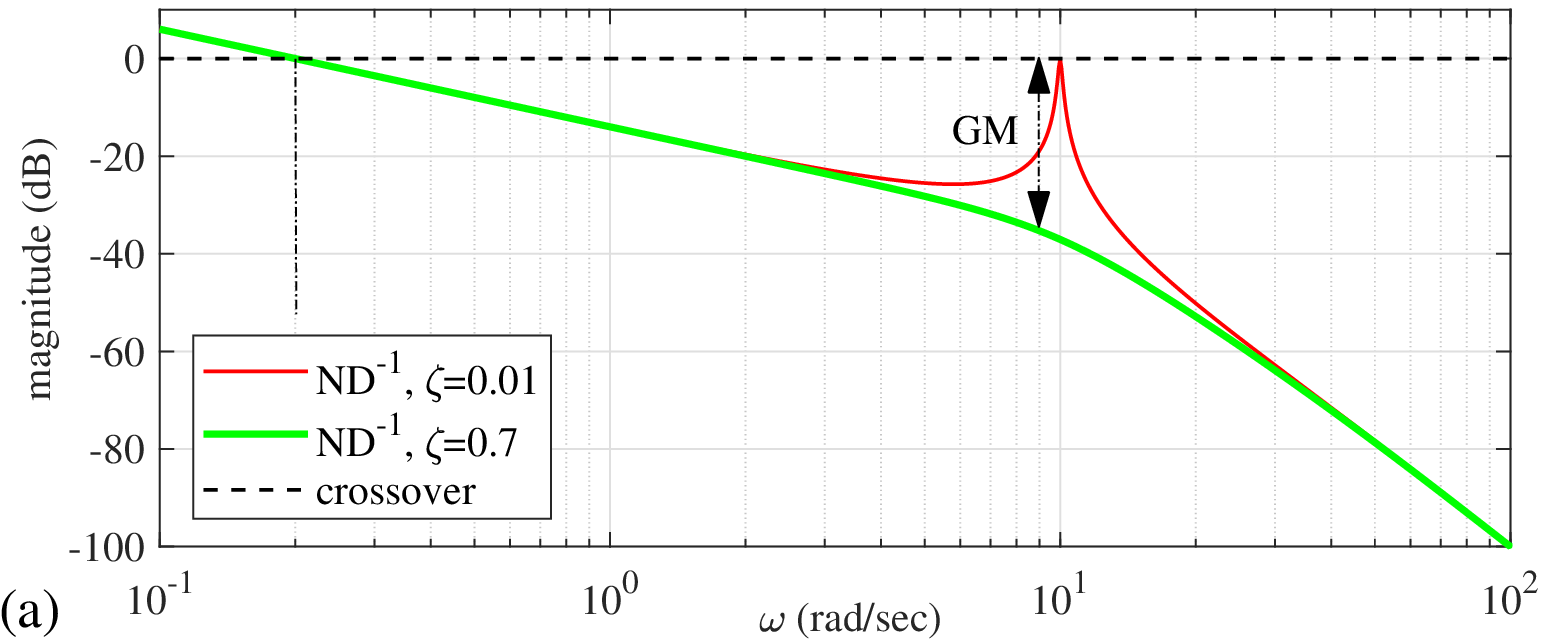}
\includegraphics[width=0.99\columnwidth]{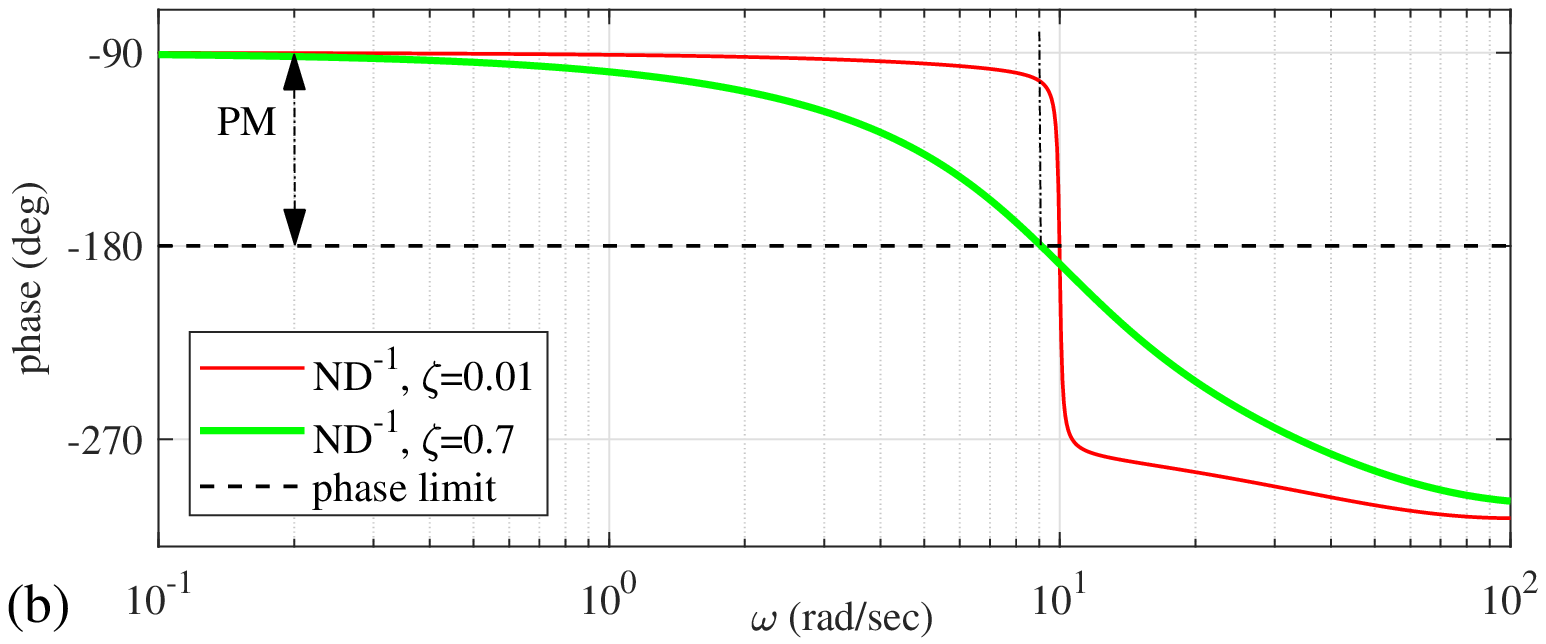}
\caption{Exemplary Bode diagrams of the loop transfer functions
$N(s)D(s)^{-1}$ with different damping ratios $\zeta = [0.01, 0.7
]$.} \label{fig:2:2}
\end{figure}

Applying any type of the output feedback controller $R(s)$, yet
without pole-zero cancelation within $R(s)G(s)$, results in the
closed-loop transfer function
\begin{equation}\label{eq:2:4}
H(s) = \frac{N(s)}{D(s)+R(s)N(s)} = \frac{N(s)D(s)^{-1}}{1 +
R(s)N(s)D(s)^{-1}}.
\end{equation}
Recall that the loop transfer function $R(s)N(s)D(s)^{-1}$
contains all information about the closed-loop behavior and is
(usually) used for analysis and design of the feedback control
$R(s)$, which should render a desired $H(s)$-behavior, cf. e.g.
\cite{skogestad2005,doyle2009}. Due to the phase lag of the
fourth-order dynamics \eqref{eq:2:3}, the output feedback
capacities are relatively limited, featured by the stability
margins, and more specifically -- gain and phase margins. An
increase of the loop gaining factor, i.e. either of $k$ in
\eqref{eq:2:3} or $R(0)$ of the control, leads unavoidably to
destabilization of $H(s)$. This is due to the lack of the gain
margin (GM) of the loop transfer function $R(s) N(s)D(s)^{-1}$
once a proportional feedback is included in $R(s)$. Worth
emphasizing is that without compensating explicitly for the
resonance peak at $\omega_0$ and, therefore, improving the GM
criteria of the loop transfer characteristics, any type of $R(s)$
with proportional feedback action will lead to an unstable, or at
least largely oscillating, behavior of $H(s)$. This becomes best
way visible when comparing the loop transfer functions
$N(s)D(s)^{-1}$ for two largely differing damping ratios, e.g.
$\zeta = [0.01, 0.7 ]$. This results in a frequency response with
and without resonance peak as exemplary shown in Figure
\ref{fig:2:2}.
While the phase margin (PM) appears sufficient and nearly
unsensitive to the damping ratio, for the assigned loop gaining
factor, the GM is drastically reduced in case of the resonance
peak. The GM of the loop transfer function with $\zeta = 0.7$,
marked by the double-arrow in Figure \ref{fig:2:2}, will only
gradually decrease with an increasing gain factor. Quite opposite,
for the loop transfer function with $\zeta = 0.01$, an
infinitesimally small gain enhancement will already lead to losing
completely GM and, therefore, to destabilization of $H(s)$.

Against the background provided above, the objective is to design
the time-delay-based output feedback compensator $R(s)$, which
will largely attenuate the resonance peak of the system
\eqref{eq:2:3} without significantly affecting the residual shape
of the loop transfer function. This way, a time-delay-based
compensation should allow for using other outer feedback control
loops which can guarantee for $x(t)-r(t) \rightarrow 0$, where $r
= \mathrm{const} \neq 0$ is a reference set value, and that after
possibly oscillation-free transient response of the control
system.

\section{Time-delay feedback control}
\label{sec:3}

The proposed control \eqref{eq:1:1} is parameterized by the gain
factor $K_d > 0$ and the time-delay
\begin{equation}
\tau = -\bigl[\arg G(j\omega) |_{\omega=\omega_0}  \bigr]
\omega_0^{-1}. \label{eq:3:1}
\end{equation}
The latter corresponds to the phase angle at the resonance
frequency, where the output value is in anti-phase (i.e. $-\pi$
rad/s) to the input value. Though an explicit use of the time
delay in output feedback was proposed previously for the
second-order systems in \cite{abdallah1993}, the control
\eqref{eq:1:1}, \eqref{eq:3:1} is principally differing to that
one provided in \cite{abdallah1993}. The difference between the
current and anti-phase-delayed output values in \eqref{eq:1:1}
provides a nearly zero or some low-constant control action for all
angular frequencies other than $\omega_0$. On the contrary, in
vicinity to $\omega_0$ the proposed control aims to suppress the
resonance peak, cf. Figure \ref{fig:2:2}, through the positive
feedback of the control law \eqref{eq:1:1}.

Despite the dynamic behavior of systems with time delay are
usually analyzed in time domain, cf. \cite{richard2003},
\cite{niculescu2004}, \cite{fridman2016}, we purposefully use the
frequency domain consideration of the system transfer
characteristics, this way allowing for stability margin analysis
and corresponding loop shaping via time delay in feedback. Here it
is worth mentioning that the stability of time-delay systems by
considering the induced-gain of a time-delay operator in the loop
and the associated Bode plots were successfully analyzed in
\cite{kao2004}.

Taking the ratio between the uncompensated loop transfer function
\eqref{eq:2:3} and feedback-compensated \eqref{eq:2:4} results in
\begin{equation}
\frac{G(s)}{H(s)} = 1 + \frac{R(s)N(s)}{D(s)}. \label{eq:3:2}
\end{equation}
For achieving the above stated compensation goals, cf. with Figure
\ref{fig:2:2}, one needs to ensure that \eqref{eq:3:2} has the
magnitude response satisfying $|G(j\omega)| |H(j\omega)|^{-1} =
\mathrm{const}$ for $\omega < \omega_0$ and $|G(j\omega)|
|H(j\omega)|^{-1} = \mathrm{const} \approx 1$ for $\omega >
\omega_0$. This is in the sense of loop shaping where the
resonance peak associated with $\omega_0$ needs to be suppressed.
Recall that here, $H(j\omega)$ constitutes a resonance compensated
loop transfer function, cf. with exemplary frequency
characteristics shown in Figure \ref{fig:2:2} for the well damped
case of $\zeta=0.7$. Note that at lower frequencies $\omega <
\omega_0$, the loop gaining factor of \eqref{eq:3:2} can be
adjusted afterwards, this way settling the required crossover
frequency once the resonance peak is compensated. For angular
frequencies around $\omega_0$, the \eqref{eq:3:2} ratio must be of
the same magnitude as the resonance peak of $G(j\omega)$, meaning
its maximal possible compensation.

\begin{figure}[!h]
\centering
\includegraphics[width=0.99\columnwidth]{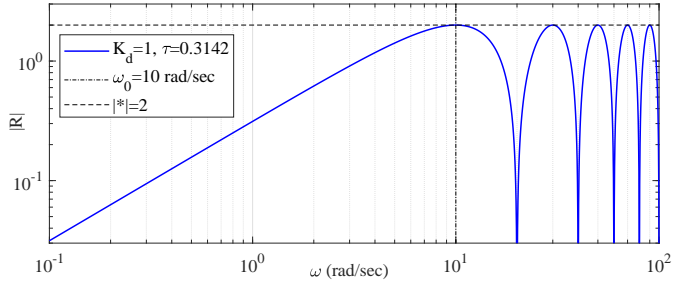}
\caption{Exemplary magnitude plot of \eqref{eq:3:3} with $K_d =1$,
$\tau=0.3142$.} \label{fig:3:1}
\end{figure}

\begin{figure}[!h]
\centering
\includegraphics[width=0.99\columnwidth]{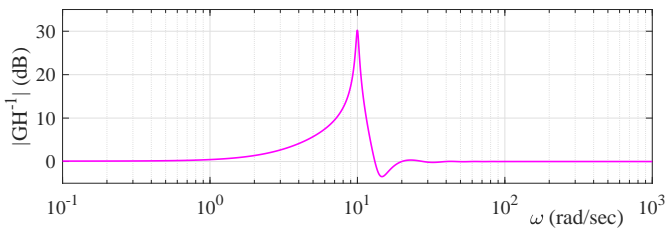}
\caption{Exemplary plot of $|GH^{-1}|$ ratio according to
\eqref{eq:3:2}, \eqref{eq:3:3}.} \label{fig:3:2}
\end{figure}

Before deriving the corresponding optimal $K_d$-gain, for which
the resonance peak is attenuated as possible at $\omega_0$, let us
first examine the principal ability of
\begin{equation}
R(s) = K_d \bigl( \exp(-s\tau) - 1 \bigr) \label{eq:3:3}
\end{equation}
to meet the above requirements at lower and higher frequencies,
i.e. for $\omega < \omega_0$ and $\omega > \omega_0$. Recall that
the time-delay-based compensator \eqref{eq:3:3}, as
infinite-dimensional operator, is upper bounded by a lead transfer
element with the gaining factor $2 K_d$, cf. Figure \ref{fig:3:1}
and e.g. \cite[ch.~4]{doyle2009}.
Since for $\omega < \omega_0$ the $|D(j\omega)|$ has also an
incremental slope of one decade per decade and $|N(j\omega)|
\rightarrow \mathrm{const} = k z_1$, it is apparent that
$|R(j\omega) N(j\omega) D^{-1}(j\omega)| = \mathrm{const}$ at
lower frequencies, cf. \eqref{eq:3:2}. Note that this constant
value depends on both, the system parameter $k z_1$ and the
control parameter $\tau$. At higher frequencies, $|D(j\omega)|$
has an incremental slope of four decades per decade, owing to four
integrators in a chain, see \eqref{eq:2:3}. At the same time,
$|N(j\omega)|$ has an incremental slope of one decade per decade
and $|R(j\omega)| \leq \mathrm{const}$ for $\omega > \omega_0$,
cf. Figure \ref{fig:3:1}. Therefore, $|R(j\omega) N(j\omega)
D^{-1}(j\omega)| \rightarrow 0$ as $\omega$ increases within the
range larger than $\omega_0$. An exemplary $G(s)H(s)^{-1}$ ratio
is demonstrated by the magnitude plot in Figure \ref{fig:3:2},
here for the sake of visualization.

In order to determine the optimal $K_d$-gain, consider first the
resonance peak of $G(j\omega)$ comparing to the transfer
characteristics of $G$ without resonance peak, i.e. for $\zeta =
0.7$. For the natural frequency, one obtains
\begin{equation}
\frac{|G(j\omega_0)|}{|G_{\zeta = 0.7}(j\omega_0)|} =
\frac{0.7}{\zeta}, \label{eq:3:4}
\end{equation}
which indicates the desired ratio of \eqref{eq:3:2} at
$\omega=\omega_0$ when applying a suitable $R(s)$. Replacing in
\eqref{eq:3:2} the left-hand-side by \eqref{eq:3:4} and $R(s)$ by
\eqref{eq:3:3}, substituting $j\omega$ instead of $s$, and
solving the obtained equation for $\omega = \omega_0$ results in
\begin{equation}
K_d = \frac{\omega_0^3 (0.7-\zeta)}{k} \, \Bigl| \frac{p_1 +
j\omega_0} { z_1 + j\omega_0 } \Bigr|. \label{eq:3:5}
\end{equation}

\section{Stabilization of non-collocated PI-control}
\label{sec:4}

For non-collocated output of the system \eqref{eq:2:3} to be
controlled, an inherent stability problem lies in the fact of a
phase 'deficit' due to the system relative degree $ > 2$. The
$\angle G$ is always crossing the $-180$ deg phase limit, cf.
Figure \ref{fig:2:2}, thus making the GM to a sensitive stability
criteria. Needless to say is that additional unmodeled lag
properties in the loop with system $G(s)$, such as due to even
minor dynamics of sensing and actuating elements and signal
transmission delays, will further increase the phase 'deficit'
and, thus, impair stability margins in general.

In order to guarantee the controlled output $x(t)$ can follow the
reference value $r(t)$, i.e. to solve not only a set value
stabilization problem $r = \mathrm{const}$, an integral control
term is usually required. When applying a standard PI
(proportional-integral) controller, with the design parameters
$K_p, K_i > 0$, the loop transfer function is written
\begin{equation}
L(s) = C(s) G(s) = K_p \, \frac{s + K_i K_p^{-1}}{s} \,
\frac{N(s)}{D(s)}. \label{eq:4:1}
\end{equation}
It is remarkable that independent of the assigned $K_p, K_i$ a
conjugate-complex pole pair of $L(s)$ does not vanish, cf. Figure
\ref{fig:2:1}, and is migrating to the right towards the unstable
right-hand-side half plane when increasing the loop gain factor
$K_p$. Using, for example, the root locus analysis or other
conventional tools of the linear control theory, cf.
\cite{skogestad2005}, one can find the critical $\max K_p$ beyond
which the closed loop of \eqref{eq:4:1} becomes unstable. Even if
$K_p < \max K_p$ is guaranteed, the uncertainties in $k$ or its
temporal variations, cf. \eqref{eq:2:3}, can destabilize the
closed-loop system. Furthermore, a lower loop gain cannot improve
the transient oscillating behavior since the damping ratio of the
conjugate-complex pole pair is not directly affected by the design
of $K_p, K_i$.

\begin{figure}[!h]
\centering
\includegraphics[width=0.99\columnwidth]{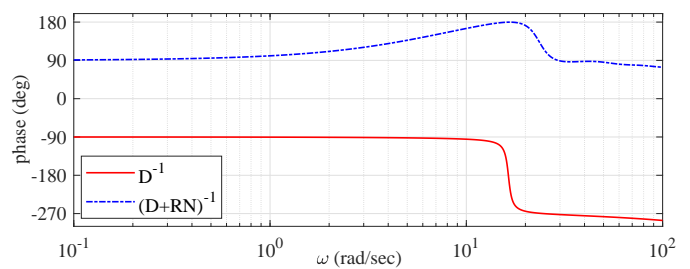}
\caption{Exemplary phase characteristics of $D^{-1}$ and $(D +
RN)^{-1}$.} \label{fig:4:1}
\end{figure}

In order to see the stabilizing properties of the time-delay-based
compensator \eqref{eq:3:3}, the $G(s)$ and $D(s)$ must be
substituted in \eqref{eq:4:1} by $H(s)$ and $D(s)+R(s)N(s)$,
respectively. Then, the phase characteristics and, thereupon
based, stability margins are visible when comparing $\angle \,
\bigl( D(j\omega) \bigr)^{-1}$ and $\angle \, \bigl( D(j\omega) +
R(j\omega)N(j\omega) \bigr)^{-1}$, see Figure \ref{fig:4:1}.
One can recognize that the lifted and reshaped phase response of
the loop transfer function with $H(j\omega)$ does not cross
$\pm180$ deg phase limits over the whole frequency range. This
allows for larger variations of the loop gains $K_p$ and $k$, in
addition to the resonance peak cancelation and, thus, improvement
of GM criterion, cf. section \ref{sec:3}.

Combining the outer PI control $C(s)$ and the time-delay-based
compensation $R(s)$ results in the overall two-degrees-of-freedom
feedback controller
\begin{equation}
u(s) = C(s)r(s) + \bigl( R(s) - C(s) \bigr) x(s). \label{eq:4:2}
\end{equation}
\begin{figure}[!h]
\centering
\includegraphics[width=0.7\columnwidth]{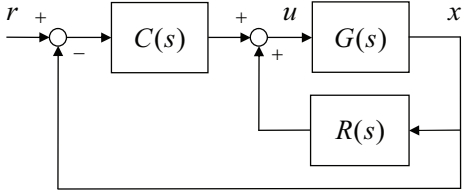}
\caption{Block diagram of the control loop with feedback
controller $C(s)$ and stabilizing compensator $R(s)$}
\label{fig:4:2}
\end{figure}
This is purposefully denoted to have two degrees-of-freedom since
the compensator $R(s)$ is designed fully independently of $C(s)$.
This is equivalent to the outer control of the resonance
compensated system $H(s)=x(s)/v(s)$, where the new (virtual) input
is $v(s) = u(s) - R(s)y(s)$. The structure of the overall
two-degrees-of-freedom control loop with \eqref{eq:4:2} is
visualized in the block diagram in Figure \ref{fig:4:2}, for
convenience of the reader.

\section{Experimental example}
\label{sec:5}

\begin{figure}[!h]
\centering
\includegraphics[width=0.6\columnwidth]{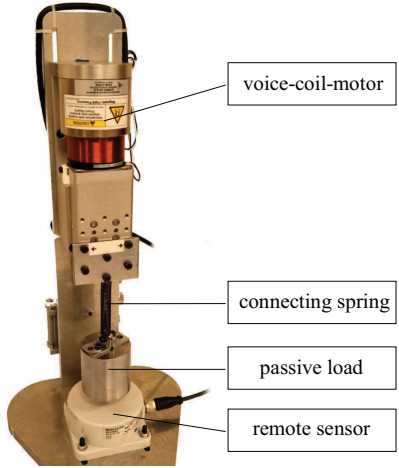}
\caption{Two-mass oscillator system with non-collocated
contactless sensing and voice-coil-motor actuation (laboratory
view).} \label{fig:5:1}
\end{figure}

\begin{figure}[!h]
\centering
\includegraphics[width=0.97\columnwidth]{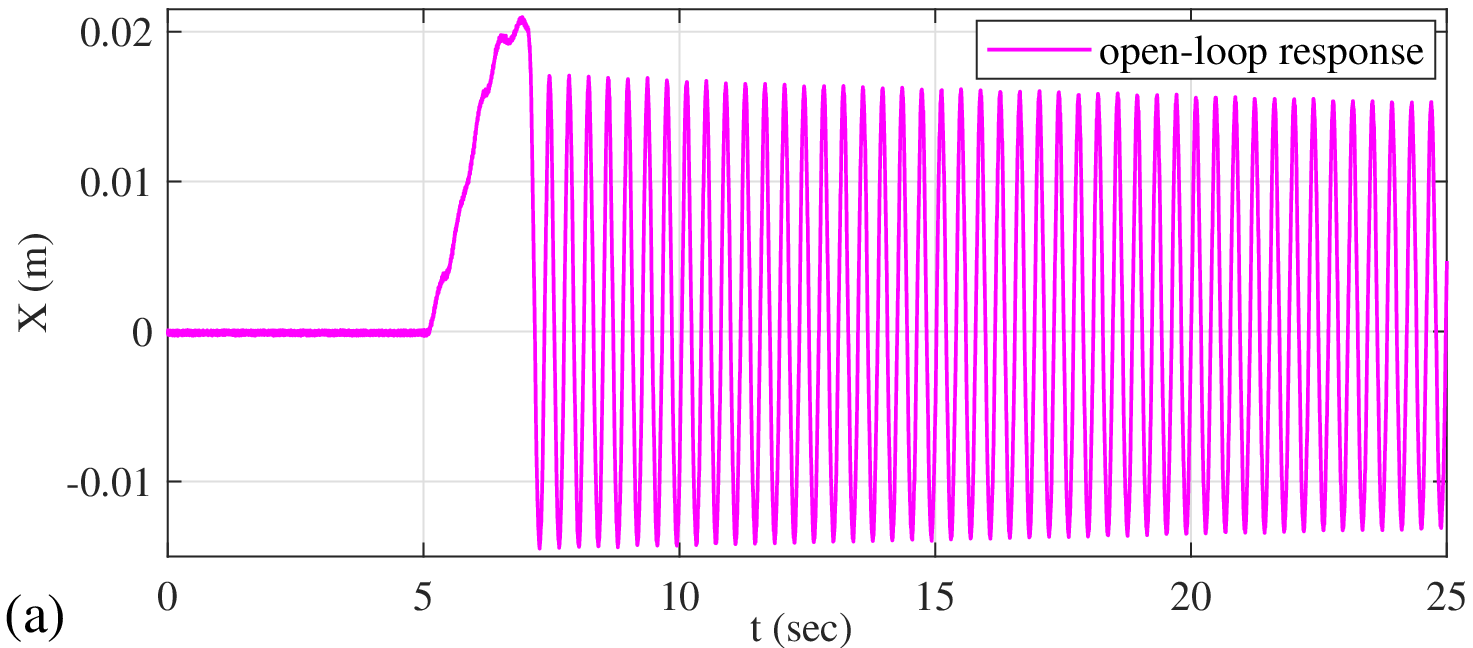}
\includegraphics[width=0.97\columnwidth]{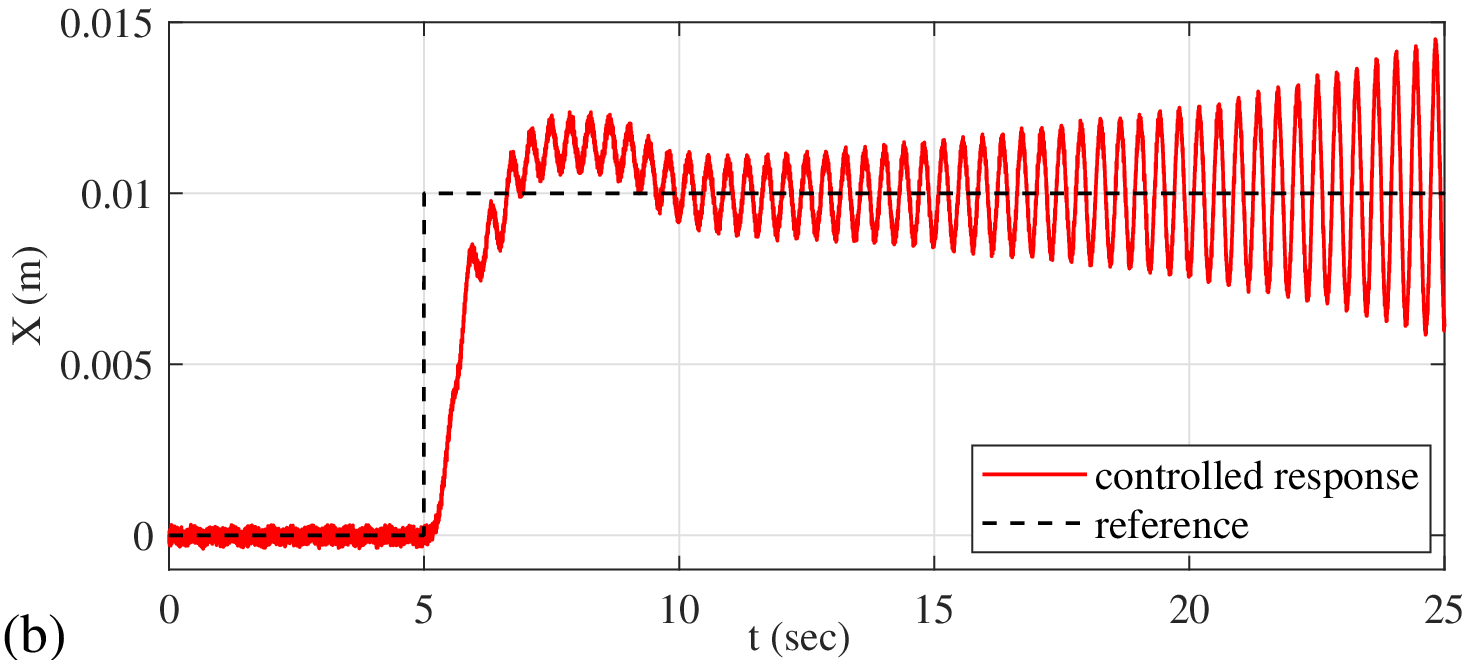}
\includegraphics[width=0.97\columnwidth]{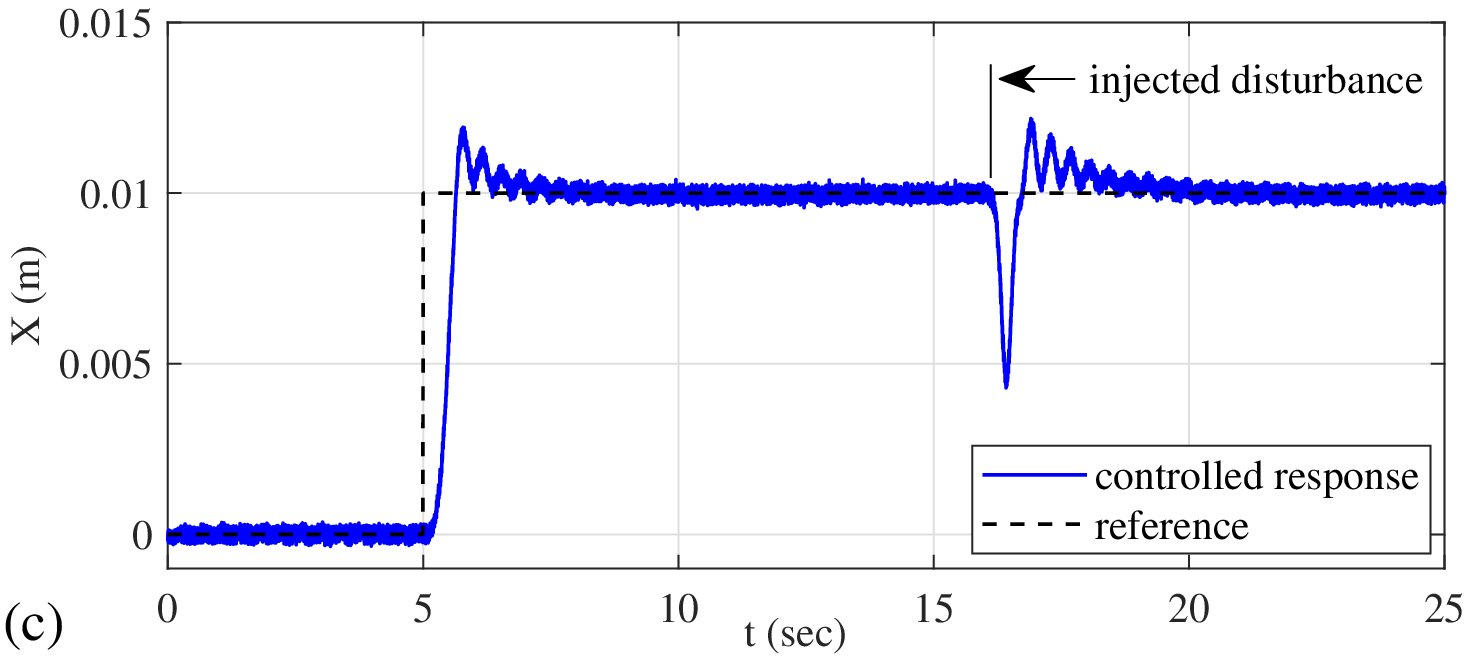}
\caption{Measured load position: oscillatory open-loop response to
a short square-shape pulse excitation in (a), controlled step
response with PI feedback only in (b), controlled step response
with PI feedback extended by the time-delay compensator
\eqref{eq:4:2} in (c).} \label{fig:5:3}
\end{figure}

In the following, a series of control experiments performed on the
laboratory system \cite{ruderman2021}, \cite{ruderman2022} is
provided for evaluating the time-delay-based compensator in accord
with sections \ref{sec:3} and \ref{sec:4}. The fourth-order system
plant is the two-mass oscillator with non-collocated contactless
sensing of the load position and actuation by the
voice-coil-motor, see Figure \ref{fig:5:1}.
The available actuator displacement is bounded by $0 \leq y \leq
0.02$ m. The well-balanced free hanging load (with one vertical
degree of freedom) is subject to the oscillations with an
extremely low structural damping of the connecting spring, cf.
\cite{ruderman2021} and Figure \ref{fig:5:3} (a). The input
control voltage $U(t)$ of the voice-coil-motor and the relative
position $X(t) = x(t) + x_0$ of the hanging load are real-time
available with the set sampling rate of 10 kHz. The steady-state
elongation offset $x_0$ is due to the gravity force when $U(t) =
0$. The nominal values of the system parameters, partially
identified and partially taken over from the technical data
sheets, are listed in Table \ref{tab:1}, while for more details on
the system model we refer to \cite{voss2022}. Since the gravity
force of both moving masses $\mathrm{m}_1$ and $\mathrm{m}_2$ is
known and does not change over the operation range, it is
pre-compensated, thus resulting in
\begin{equation}\label{eq:5:1}
U(t) = u(t) + u_g = u(t) + \frac{\mathrm{R} \mathrm{g}}{\Psi}
\Bigl(\mathrm{m}_1+\mathrm{m}_2\Bigr).
\end{equation}
Here $u(t)$ is the applied feedback control law \eqref{eq:4:2}.
Further we note that the electromagnetic dynamics of the
voice-coil-motor is reasonably neglected, so that the coupling
factor between the input (terminal) voltage and the produced
electro-magnetic force (EMF) is $\Psi \mathrm{R}^{-1}$.
\begin{table}[!h]
  \renewcommand{\arraystretch}{1.3}
  \caption{Nominal values of the system parameters.}
  \footnotesize
  \label{tab:1}
  \begin{center}
  \begin{tabular} {|p{1.3cm}|p{1cm}|p{1cm}|p{3.5cm}|}
  \hline \hline
  parameter &  unit   &  value & meaning        \\
  \hline \hline
  $\mathrm{m}_1$ &    kg    &   0.6      &  actuator mass  \\
  \hline
  $\mathrm{m}_2$ &    kg    &   0.75     &  load mass  \\
  \hline
  $\mathrm{k}$   &    N/m   &    200     &  spring constant  \\
  \hline
  $\sigma$   &    kg/s   &    200     &  actuator damping  \\
  \hline
  $\delta$   &    kg/s   &    0.01     &  spring damping  \\
  \hline
  $\mathrm{R}$        &    V/A    &    5.23     &  coil resistance  \\
  \hline
  $\Psi$     &    Vs/m   &    17.16    &  EMF constant  \\
  \hline
  $\mathrm{g}$   &    m/s$^2$     &    9.81     &  gravity constant  \\
  \hline \hline
  \end{tabular}
  \end{center}
  \normalsize
\end{table}

\begin{figure}[!h]
\centering
\includegraphics[width=0.99\columnwidth]{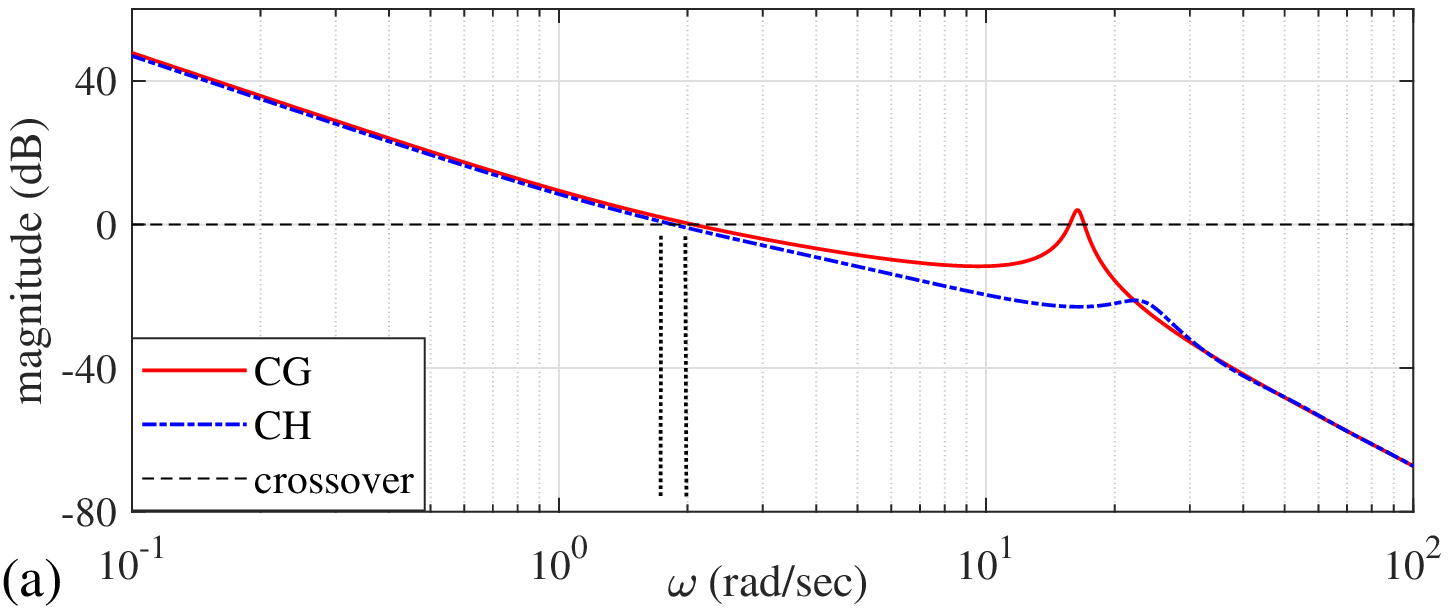}
\includegraphics[width=0.99\columnwidth]{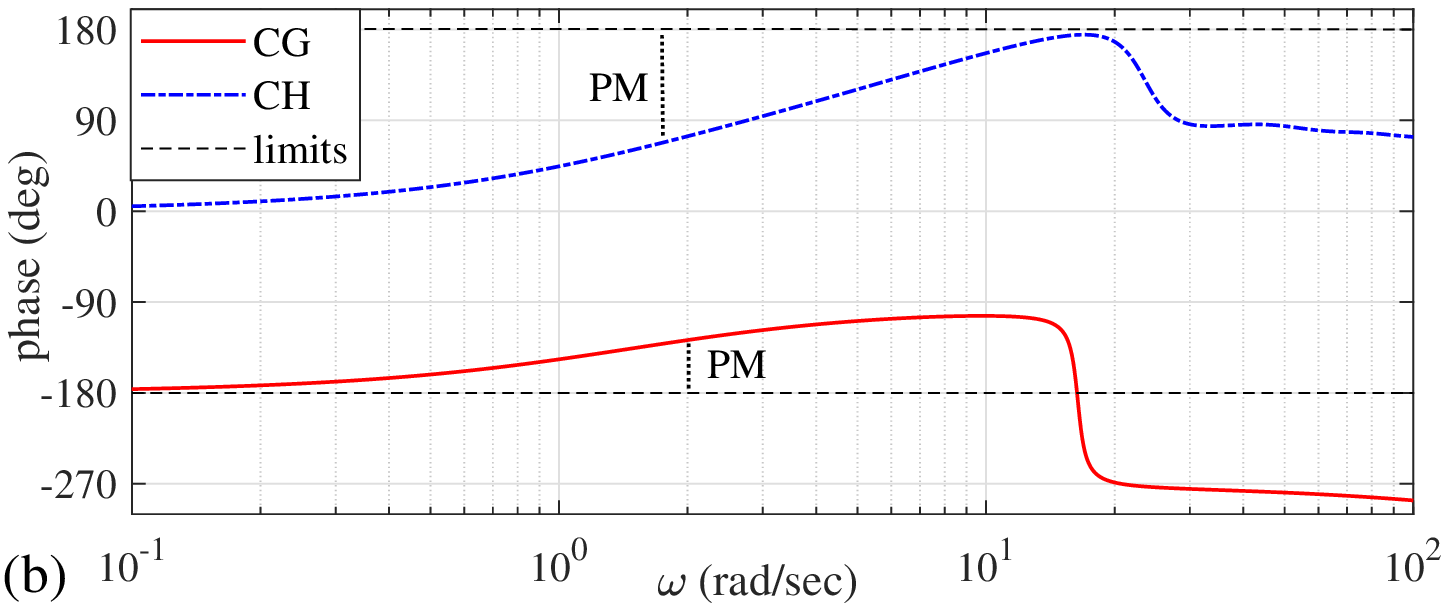}
\caption{Bode diagrams of the loop transfer functions $C G$ and $C
H$.} \label{fig:5:2}
\end{figure}

For the vector of the state variables $z \equiv (\dot{y}, y,
\dot{x}, x )^\top$, the corresponding state-space model is
\begin{equation}\label{eq:5:2}
\dot{z} = \underset{A} {\underbrace{
\left(%
\begin{array}{cccc}
-333.4 &  -333.3  &    0.033      & 333.3  \\
1         & 0     &     0         & 0         \\
0.027     & 266.7 &    -0.027     & -266.7 \\
0         & 0          &  1       & 0
\end{array}%
\right) } } z + B \,u,
\end{equation}
with the coupling vectors $B = (5.47, 0, 0, 0)^\top$ and $F = (0,
0, 0, 1)$ of the input and output, respectively. The natural
frequency corresponding to the oscillations of the hanging load is
at $\omega_0 = 16.3$ rad/sec. The parameters of the feedback
control $u(t)$, given by \eqref{eq:4:2}, are $K_p = 100$, $K_i =
150$, $K_d = 100$, $\tau = 0.1923$. The corresponding Bode
diagrams of the plant $G(j\omega)$ and the plant extended by the
time-delayed feedback $H(j\omega)$, both connected in series with
the PI-feedback controller $C(j\omega)$, are shown in Figure
\ref{fig:5:2}.
While the phase margin of the $C(s)G(s)$ loop transfer function is
relatively moderate, being $\mathrm{PM}_{CG} = 52.3$ deg, the gain
margin is not available, i.e. being already negative
$\mathrm{GM}_{C G} = -4$ dB. That means an unstable closed-loop
behavior of the $C(s)G(s)$ loop system. On the contrary, the
$C(s)H(s)$ loop transfer function reveals a sufficiently large
phase margin, as $\mathrm{PM}_{CH} = 112$ deg, and a theoretically
infinite gain margin since $\angle CH$ is not crossing the $\pm
180$ deg phase limits, see Figure \ref{fig:5:2}.

The open-loop response of the experimental system to a short
rectangular pulse is shown in Figure \ref{fig:5:3} (a) and
visualizes the low damping properties of the two-mass oscillator.
The unstable step response of the PI-feedback controlled system
(i.e. when setting $K_d=0$) is shown in Figure \ref{fig:5:3} (b).
On the contrary, the step response of the time-delay-stabilized
PI-feedback control system (i.e. when allowing for $K_d = 100$) is
shown in Figure \ref{fig:5:3} (c). Note that here, an additional
external manual disturbance was injected at the time about $t
\approx 16.5$ sec, for additional assessment of robustness of the
proposed time-delay-based compensator.

\section{Summary and
Discussion} \label{sec:6}

In this paper, we presented a new time-delay-based control method
which allows for a robust compensation of resonance oscillations
in non-collocated fourth-order dynamic systems. The compensator
has only two parameters, the natural frequency and an adjustable
gain factor, the value of which is determined based on the
knowledge of resonance peak magnitude of the system transfer
characteristics. The analysis of the time-delay-based approach and
associated loop shaping are made in frequency domain, also using
the stability margins as classical criteria for robustness and
performance of an output feedback loop.

It is fair to notice that the proposed control \eqref{eq:1:1},
\eqref{eq:3:1} has the induced gain properties which are similar
(in magnitude response) to those of the lead transfer element
$$
W(s) = \frac{2 K_d s}{\tau s + 1}.
$$
The main advantage of the proposed method, however, is that
neither time derivatives of the noisy output nor implementation of
any transfer functions, like e.g. $W(s)$, are required for
applying \eqref{eq:1:1}, \eqref{eq:3:1}. Also the analytic form
\eqref{eq:1:1} of the control law in time domain can allow for
adaptation and on-line adjustment of the natural frequency
parameter, which speaks for different real applications.

If a non-collocated output feedback system \eqref{eq:2:3} uses a
PI control to guarantee the steady-state performance, i.e.
$x(t)-r(t) \rightarrow 0$, the feedback control loop yields
inherently unstable if not compensating for the resonance peak.
The proposed time-delay-based loop shaping suppresses robustly the
resonance peak without much affecting the loop transfer
characteristics at other frequencies. It is worth noting that the
proposed compensator relies on the knowledge of $\omega_0$, cf.
\eqref{eq:3:1}, \eqref{eq:3:5}. A robust estimation of $\omega_0$,
as shown in \cite{ruderman2022}, is however available for tuning
the required parameters. It can also be noted that despite
uncertainties of $\omega_0$, which are due to effective stiffness
in the operational point of the oscillator, the compensator
performs robustly, as confirmed by the experimental evaluation. A
detailed sensitivity analysis of parameter tuning is beyond the
scope of this work and is the subject of future research.

The proposed control method was evaluated experimentally on
two-mass oscillator system with non-collocated contactless sensing
and voice-coil-motor actuation. It was shown, cf. Figure
\ref{fig:5:3}, that an extremely low-damped oscillatory behavior
is effectively compensated by the proposed time-delay-based
method, thus allowing also for a standard PI outer feedback
control loop.

\bibliographystyle{elsarticle-num}
\bibliography{references}

\end{document}